\documentstyle[epsf]{article}

\title{Quantum phase transitions
       in alternating transverse Ising chains:
       \protect\\
       Rigorous analytical and exact numerical results}

\author{Oleg Derzhko$^{1,2}$,
        Johannes Richter$^3$,
        Taras Krokhmalskii$^1$
        and Oles' Zaburannyi$^1$\\
\small {$^1$Institute for Condensed Matter Physics,}\\
\small {1 Svientsitskii Str., L'viv-11, 79011, Ukraine}\\
\small {$^2$Chair of Theoretical Physics,
            Ivan Franko National University of L'viv,}\\
\small {12 Drahomanov Str., L'viv-5, 79005, Ukraine}\\
\small {$^3$Institut f\"{u}r Theoretische Physik,
            Universit\"{a}t Magdeburg,}\\
\small {P.O. Box 4120, D-39016 Magdeburg, Germany}}

\date{\today}

\begin{document}

\renewcommand\baselinestretch {1.3}
\large\normalsize

\maketitle

\begin{abstract}
We examine the ground state properties
of the $s=\frac{1}{2}$ transverse Ising chain
with regularly alternating bonds and fields
using
exact analytical results
and
exact numerical data
for long (up to $N=900$)
and
short ($N=20$) chains.
For a given period of alternation the system
may exhibit a series of quantum phase transitions,
where the number of
transitions depends on the concrete set of the parameters of the
Hamiltonian.
The critical behaviour
for the nonuniform and the uniform chain is the same.

\end{abstract}

\vspace{1cm}

\noindent
{\bf PACS number(s):}
75.10.-b

\vspace{5mm}


\renewcommand\baselinestretch {1.5}
\large\normalsize

Currently the investigation of quantum phase transitions
is a very active field of physical research \cite{001,002}.
Among others the quantum magnets
represent an interesting class of systems
studied theoretically and experimentally
where quantum phase transitions appear
\cite{002,003,004}.
The one-dimensional $s=\frac{1}{2}$ Ising model in a transverse field
(the transverse Ising chain)
described by the Hamiltonian
\begin{eqnarray}
\label{001}
H=\sum_n\Omega s_n^z +\sum_n 2I s_n^xs_{n+1}^x
\end{eqnarray}
($s_n^{\alpha}=\frac{1}{2}\sigma_n^{\alpha}$,
$\sigma_n^{\alpha}$ are the Pauli matrices
attached to the site $n$,
$\Omega$ is the on-site transverse field,
$2I$ is the exchange interaction between neighbouring spins)
is known to be the simplest system exhibiting a quantum
phase transition.

For positive $\Omega$
the system exhibits a zero temperature
transition from the ordered (quantum Ising) phase
(for $\Omega< \Omega_c$)
to the disordered (quantum paramagnetic) phase
(for $\Omega > \Omega_c$)
at
a critical point $\Omega_c =\vert I\vert$ \cite{002}.
(Notice,
that the same kind of transition
takes place at $\Omega_c =-\vert I\vert$.)
There has been a great deal of theoretical work
on the ground state (GS) properties
of the transverse Ising chain (\ref{001}).
The knowledge of the eigenvalues and eigenfunctions
of the Hamiltonian (\ref{001})
\cite{005,006,002}
makes the problem accessible for exact analysis.
Let us briefly summarize some of the main exact results for
the GS and the excitation gap.
(i)
At the critical transverse field $\Omega_c$
the gap in the energy spectrum $\Delta$
closes as
$\Delta\sim\vert\Omega-\Omega_c\vert$.
(ii)
The GS correlation function
$\langle s^x_j s^x_{j+n} \rangle$
tends to ${m^x}^2\ne 0$ as $n\to\infty$
for $0\le\Omega<\Omega_c$
with
the longitudinal (Ising) magnetization
$m^x=\frac{1}{2}
\left(1-\left(\frac{\Omega}{\Omega_c}\right)^2\right)^{\frac{1}{8}}$
which plays the role of the order parameter.
(iii)
For  $\Omega>\Omega_c$  the correlation function
$\langle s^x_j s^x_{j+n} \rangle$
decays exponentially to zero for large $n$.
(iv)
Close to the critical point $\Omega_c$
the transverse magnetization $m^z$
and
the static transverse susceptibility $\chi^z$
behave as
$m^z\sim \left(\Omega-\Omega_c\right)
\ln \vert\Omega-\Omega_c\vert$
and
$\chi^z\sim \ln \vert\Omega-\Omega_c\vert$,
respectively.
(v)
At the critical point $\Omega=\Omega_c$
the correlation length diverges
and $\langle s^x_j s^x_{j+n} \rangle$
becomes scale invariant,
i.e.,
it goes to zero for large $n$ according to a power-law
($\langle s^x_j s^x_{j+n} \rangle \sim n^{-\frac{1}{4}}$).
(vi)
As can be seen from the above results
the set of the critical exponents for the
quantum phase transition in dimension one
is identical to that for the thermal phase transition
of the square-lattice Ising model.

Recently several authors
(see, e.g., Refs. \cite{007,008,009,010,011,012,013,014})
have addressed  the question,
how deviations from the pure uniform crystalline system
(e.g., disorder, aperiodically or regularly varying parameters)
influence the properties of quantum spin systems.
In what follows we study this question
for the
transverse Ising model with fields and interactions
which vary regularly from site to site
with a finite period $p$,
i.e., we change in (\ref{001})
$\Omega\to\Omega_n$,
$I\to I_n$
and the sequence of parameters along the chain is
$\Omega_1I_1\Omega_2I_2\ldots\Omega_pI_p
\Omega_1I_1\Omega_2I_2\ldots\Omega_pI_p\ldots\;$.
In particular,
we examine the influence of the so-defined regular
inhomogeneity on the existence and the nature of the
quantum phase transition.
In order to study the properties
of the regularly alternating transverse Ising chain
we use
i) exact analytical results for thermodynamic quantities
obtained by means of continued fractions
\cite{015},
ii) exact analytical results for the GS wave function for special
parameter values,
iii) exact numerical data for long chains of up to $N=900$ spins
\cite{016},
and
iv) exact diagonalization data
for short chains of $N=20$ spins.

As described in \cite{015}
the thermodynamic quantities
of the regularly alternating transverse Ising chain
can be calculated rigorously
for any finite period of inhomogeneity $p$.
After using the Jordan-Wigner fermionization
the transverse Ising chain
is described by noninteracting spinless fermions
with the Hamiltonian
$H=\sum_k\Lambda_k\left(\eta_k^+\eta_k-\frac{1}{2}\right)$
and
the distribution of the square of energy of the elementary
excitations
$R(E^2)=\frac{1}{N}\sum_k\delta\left(E^2-\Lambda_k^2\right)$
can be found exactly
by means of continued fractions.
The resulting density of states has the form
\begin{eqnarray}
\label{002}
R(E^2)=
\left\{
\begin{array}{ll}
\frac{1}{p\pi}
\frac{\vert {\cal{Z}}_{p-1}(E^2)\vert}
{\sqrt{{\cal{A}}_{2p}(E^2)}}, &
{\mbox{if}}\;\;\;
{\cal{A}}_{2p}(E^2)>0, \\
0, &
{\mbox{otherwise}},
\end{array}
\right.
\end{eqnarray}
where ${\cal{Z}}_{p-1}(E^2)$
and ${\cal{A}}_{2p}(E^2)=-\prod_{j=1}^{2p}\left( E^2-a_j\right)$
are polynomials of  $(p-1)$th and $(2p)$th orders,
respectively,
and the $a_j \ge 0$ are the roots of ${\cal{A}}_{2p}(E^2)$.
For example,
${\cal{Z}}_{1}(E^2)
=2E^2-\Omega_1^2-\Omega_2^2-I_1^2-I_2^2$,
${\cal{A}}_{4}(E^2)
=4\Omega_1^2\Omega_2^2I_1^2I_2^2
-\left(E^4-\left(\Omega_1^2+\Omega_2^2+I_1^2+I_2^2\right)E^2
+\Omega_1^2\Omega_2^2
+I_1^2I_2^2\right)^2$
(for details of calculation see \cite{015}).
$R(E^2)$ permits to obtain the energy gap $\Delta$
for the spin model,
since the smallest root of ${\cal{A}}_{2p}(E^2)$
yields the squared minimal excitation energy of the spin system.
Knowing $R(E^2)$
we immediately get all GS quantities
which follow from the GS energy per site
$e_0=-\int_0^{\infty}{\mbox{d}}E E^2 R(E^2)$.
Assuming that $\Omega_n=\Omega+\Delta\Omega_n$
we obtain the GS transverse magnetization
$m^z=\frac{\partial e_0}{\partial \Omega}$
and the GS static transverse susceptibility
$\chi^{z}=\frac{\partial m^z}{\partial \Omega}$.
The thermodynamic quantities at nonzero temperature
follow from the Helmholtz free energy per site
$f=-2kT\int_0^{\infty}{\mbox{d}}E E R(E^2)
\ln\left(2\cosh\frac{E}{2kT}\right)$.

Within the analytical approach based on continued fractions
it is not possible to calculate the spin correlation functions.
However, we can calculate
the GS spin correlation functions
$\langle s_j^{\alpha} s_{j+n}^{\alpha} \rangle$ ($\alpha=x, z$)
numerically for finite chains.
It is a specific feature of the considered quantum spin model
that the $2^{N}\times 2^{N}$ problem
inherent in the exact diagonalization scheme
reduces to a $N\times N$ problem
\cite{006,016,002}
and therefore numerical exact data for rather
long chains can be obtained.
(Notice, that typical chain length
accessible for Lanczos diagonalization of quantum Heisenberg
systems with low  symmetry  is about $N=24$.)
The on-site magnetizations can be calculated from the
correlation functions
using the relation
$m^{\alpha}_{j_1}m^{\alpha}_{j_2}
={\mbox{lim}}_{r\to\infty}
\langle s_{j_1}^{\alpha} s_{j_2+rp}^{\alpha} \rangle$,
where $1\ll j_1\ll N$ is one of the $p$ consecutive numbers
sufficiently far from the ends of chain
and $j_2-j_1=0,1,\ldots,p-1$.
Assuming that
$\langle s_j^{\alpha} s_{j+rp}^{\alpha} \rangle
-\langle s_j^{\alpha} \rangle \langle s_{j+rp}^{\alpha} \rangle
\sim\exp\left(-\frac{rp}{\xi^{\alpha}}\right)$
for large $r$
($1\ll j,j+rp\ll N$)
we can determine the correlation length $\xi^{\alpha}$ from the
exact numerical data.

Let us discuss the effects of regular alternation
on the existence, the position and the critical properties
of the quantum phase transition
in the transverse Ising chain in more detail.
For a certain set of parameters
we identify the points of quantum phase transition
by looking for vanishing excitation gap $\Delta$.
As mentioned above
$\Delta$ is determined
by the smallest root of ${\cal{A}}_{2p}(E^2)$
and therefore the location of a quantum phase transition
is determined  by
${\cal{A}}_{2p}(0)=0$.
For period $p=2$ the corresponding condition
${\cal{A}}_4(0)=0$
yields
\begin{eqnarray}
\label{003}
\Omega_1\Omega_2=\pm I_1I_2.
\end{eqnarray}
It should be mentioned
that a condition for closing of the excitation gap
for an inhomogeneous transverse Ising chain
was already reported in Ref. \cite{017}.
Eq. (\ref{003}) agrees with that result
for the regularly alternating chain of period 2.
In what follows we consider as an example
a chain with
$\Omega_{1,2}=\Omega\pm\Delta\Omega$, $\Delta\Omega\ge 0$.
We find either two characteristic fields
$\Omega_c=\pm\sqrt{{\Delta\Omega}^2+\vert I_1 I_2\vert}$
if
$\Delta\Omega<\sqrt{\vert I_1 I_2\vert}$
or
four characteristic fields
$\Omega_c=\pm\sqrt{{\Delta\Omega}^2\pm \vert I_1 I_2\vert}$
if
$\Delta\Omega>\sqrt{\vert I_1 I_2\vert}$ \cite{018}.
This is illustrated in Figs. 1a, 1d
where the gap
$\Delta$ in dependence on $\Omega$ (dotted curves)
for the chains with $\vert I_1\vert=\vert I_2\vert=1$
and
$\Delta\Omega=0.5$ (Fig. 1a)
and
$\Delta\Omega=1.5$ (Fig. 1d)
is shown.
Thus,
in case of small strength of transverse-field inhomogeneity $\Delta\Omega$
only quantitative deviations from the homogeneous case may be expected.
Contrary,
strong inhomogeneity $\Delta\Omega$ increases the number of quantum phase transitions
\cite{019}.
Using the explicit expression for ${\cal{A}}_{4}(E^2)$
we find
that the smallest root $a_j$
decays as
$\sim\left(\Omega-\Omega_c\right)^2$
for
$\Omega\to\Omega_c$.
Therefore,
the energy gap $\Delta$ closes as
$\sim\vert\Omega-\Omega_c\vert$
for $\Omega\to\Omega_c$, whereas
the GS energy $e_0$
contains the term
$\sim\left(\Omega-\Omega_c\right)^2\ln\vert\Omega-\Omega_c\vert$
and hence
$m^z$ and $\chi^z$
contain the terms
$\sim\left(\Omega-\Omega_c\right)\ln\vert\Omega-\Omega_c\vert$
and
$\sim\ln\vert\Omega-\Omega_c\vert$,
respectively.
This means that the critical behaviour 
of the regularly alternating transverse Ising chain
is identical to that of the homogeneous chain.
In Figs. 1a, 1d we show
$m^z$ vs. $\Omega$ (solid curves) and
$\chi^z$ vs. $\Omega$ (dashed curves)
for $\Delta\Omega=0.5$ (Fig. 1a)
and
$\Delta\Omega=1.5$ (Fig. 1d).
We emphasize,
that the quantum phase transition is not suppressed
by the deviation of the uniformity
in kind of a regular alternation of bonds and fields.
Even the appearance of additional transitions
due to field alternation is possible.

In addition to the exact results for thermodynamic quantities
we can obtain exact expressions
for the GS $\vert{\mbox{GS}}\rangle$
of the considered chain at special values of $\Omega$.
Obviously, we have
$\vert {\mbox{GS}} \rangle
=\prod_n\vert \uparrow_n\rangle$
($\vert {\mbox{GS}} \rangle
=\prod_n\vert \downarrow_n\rangle$)
for $\Omega\to -\infty$ ($\Omega\to +\infty$),
$\vert {\mbox{GS}} \rangle
=\ldots\vert\downarrow_n\rangle\vert\uparrow_{n+1}\rangle\ldots\;$
for $\Omega=0$
and
$\Delta\Omega\gg \vert I_1\vert,\vert I_2\vert$ and
$\vert {\mbox{GS}} \rangle
=\prod_n
\frac{1}{\sqrt{2}}
\left(\vert\downarrow_n\rangle+\vert\uparrow_n\rangle\right)$
or
$\vert {\mbox{GS}} \rangle
=\prod_n
\frac{1}{\sqrt{2}}
\left(\vert\downarrow_n\rangle-\vert\uparrow_n\rangle\right)$
for $\Omega=\Delta\Omega=0$
(we have assumed
the ferromagnetic sign of the exchange interactions).
Nontrivial but solvable is the case
$\Omega=\mp\Delta\Omega$,
i.e.,
$\Omega_1=0$, $\Omega_2=-2\Delta\Omega$
or
$\Omega_1=2\Delta\Omega$, $\Omega_2=0$.
Supposing $I_1I_2>0$
the GS
for $\Omega_n=0$, $\Omega_{n+1}=-2\Delta\Omega$
is given by
\begin{eqnarray}
\label{004}
\vert {\mbox{GS}} \rangle
=U^+R^+\ldots
\left(c_1\vert\downarrow_{n+1}\rangle\vert\downarrow_{n+2}\rangle
+c_2\vert\uparrow_{n+1}\rangle\vert\uparrow_{n+2}\rangle\right)
\left(c_1\vert\downarrow_{n+3}\rangle\vert\downarrow_{n+4}\rangle
+c_2\vert\uparrow_{n+3}\rangle\vert\uparrow_{n+4}\rangle\right)
\ldots,
\\
U^+=\ldots\exp{\left(-{\mbox{i}}\pi s_n^xs_{n+1}^y\right)}
\exp{\left(-{\mbox{i}}\pi s_{n-1}^xs_n^y\right)}\ldots,
\nonumber\\
R^+=\ldots\exp{\left(-{\mbox{i}}\frac{\pi}{2}s_n^z\right)}
\exp{\left(-{\mbox{i}}\frac{\pi}{2}s^z_{n-1}\right)}\ldots,
\nonumber\\
c_1=\frac{1}{\sqrt{2}}\frac{I+\sqrt{I^2+\Delta\Omega^2}}
{\sqrt{I^2+I\sqrt{I^2+\Delta\Omega^2}+\Delta\Omega^2}},
\;\;\;
c_2=\frac{1}{\sqrt{2}}\frac{\Delta\Omega}
{\sqrt{I^2+I\sqrt{I^2+\Delta\Omega^2}+\Delta\Omega^2}},
\;\;\;
I=\frac{I_1+I_2}{2}.
\nonumber
\end{eqnarray}
To obtain (\ref{004}) we note
that the transformed Hamiltonian $RUHU^+R^+$ for $N\to\infty$
(when boundary terms become negligible)
describes a system of noninteracting two-site clusters
each with the Hamiltonian
$I_1s_{n+1}^z+I_2s_{n+2}^z
- 4\Delta\Omega s_{n+1}^xs_{n+2}^x$.
For $\Omega_n=2\Delta\Omega$, $\Omega_{n+1}=0$
the GS is again given by (\ref{004})
with the change $n\to n-1$, $\Delta\Omega\to-\Delta\Omega$.
Knowing  $\vert {\mbox{GS}} \rangle$
we can calculate the $xx$ and $zz$ spin correlation functions.
For example,
for $\Omega=-\Delta\Omega$
according to Eq. (\ref{004}) we find
\begin{eqnarray}
\label{005}
\langle s_{n+1}^x s_{n+2}^x \rangle
=\langle s_{n+1}^x s_{n+4}^x \rangle
=\ldots
=\langle s_{n+2}^x s_{n+3}^x \rangle
=\langle s_{n+2}^x s_{n+5}^x \rangle
=\ldots
=\frac{1}{4}\left(c_2^2-c_1^2\right),
\nonumber\\
\langle s_{n+1}^x s_{n+3}^x \rangle
=\langle s_{n+1}^x s_{n+5}^x \rangle
=\ldots
=\frac{1}{4}\left(c_2^2-c_1^2\right)^2,
\;\;\;
\langle s_{n+2}^x s_{n+4}^x \rangle
=\langle s_{n+2}^x s_{n+6}^x \rangle
=\ldots
=\frac{1}{4};
\nonumber\\
\langle s_{n+1}^z s_{n+2}^z \rangle
=\langle s_{n+1}^z s_{n+4}^z \rangle
=\ldots
=\langle s_{n+2}^z s_{n+3}^z \rangle
=\langle s_{n+2}^z s_{n+5}^z \rangle
=\ldots
=\langle s_{n+2}^z s_{n+4}^z \rangle
=\langle s_{n+2}^z s_{n+6}^z \rangle
=\ldots
=0,
\nonumber\\
\langle s_{n+1}^z s_{n+3}^z \rangle
=\langle s_{n+1}^z s_{n+5}^z \rangle
=\ldots
=\left(c_1c_2\right)^2.
\end{eqnarray}
For $\Omega=\Delta\Omega$
the correlation functions are given by (\ref{005})
with the change $n\to n-1$.

Now we discuss the exact numerical results for finite chains
presented in Fig. 1 and Fig. 2.
The order parameters (longitudinal sublattice magnetizations)
$\vert m^x_j\vert$ versus transverse field $\Omega$
are  shown in Figs. 1 and 2 for
$\Delta\Omega<\sqrt{\vert I_1I_2\vert}$
(Figs. 1b, 2a (open symbols))
and
$\Delta\Omega>\sqrt{\vert I_1I_2\vert}$
(Figs. 1e, 2a (full symbols)).
In accordance with the analytical results for $\Delta$, $m^z$
and $\chi^z$
the numerical data show the existence of
either two phases
(quantum Ising phase for
$\vert \Omega\vert <\sqrt{{\Delta\Omega}^2+\vert I_1 I_2\vert}$
and strong-field quantum paramagnetic phase otherwise)
or three phases
(low-field quantum paramagnetic phase for
$\vert \Omega\vert <\sqrt{{\Delta\Omega}^2-\vert I_1 I_2\vert}$,
quantum Ising phase for
$\sqrt{{\Delta\Omega}^2-\vert I_1 I_2\vert}
<\vert \Omega\vert
<\sqrt{{\Delta\Omega}^2+\vert I_1 I_2\vert}$
and strong-field quantum paramagnetic phase otherwise).
The longitudinal on-site magnetizations $m_j^x$ 
are nonzero 
in quantum Ising phases 
and become strictly zero 
in quantum paramagnetic phases.
The transverse magnetization $m^z$ in quantum paramagnetic phases  
is almost constant 
being in the vicinity 
of zero in the low-field phase 
and 
of saturation value in the strong-field phase
thus producing plateau-like steps in the dependence $m^z$ versus $\Omega$.
The  correlation length $\xi^x$
(Figs. 1c, 1f, 2b (full symbols))
illustrates that the transition between different phases
is accompanied by the divergency of $\xi^x$.
As can be seen in Figs. 1b, 1e
short chains are sufficient to
reproduce the correct order-parameter behaviour
in the quantum Ising phase away from the quantum critical point,
but not in the quantum paramagnetic phases.
The chain length of up to
$N=900$ sites is clearly sufficient to see a sharp transition in the order
parameter (Figs. 1b, 1e) and even to extract reliable results for $\xi^x$
from the long-distance  behaviour of $\langle s_j^{x} s_{j+r}^{x} \rangle$
as can be seen from the data  presented in Figs. 1c, 1f and in Fig. 2b.
We note that the numerical data for
$\vert m_j^x\vert$ and $\xi^x$ at $\Omega=\mp\Delta\Omega$
coincide with the exact expression (\ref{005}).
In particular,
the values of the longitudinal sublattice magnetizations
for these fields
are
$\frac{1}{2}$
and
$\frac{1}{2}\vert c_2^2-c_1^2\vert$.
It should be remarked that inhomogeneity inherent in the model 
is regular but not random 
and there is no frustration;
as a result we do not observe any indication of a glassy behaviour.

Finally,
the low-temperature dependence of the specific heat $c$
at different $\Omega$ ($\ge 0$)
obtained from the exact analytical expression for the free energy (see above)
confirms the
existence of
either two phase transitions
(Fig. 3a)
or four phase transitions
(Fig. 3b)
depending on the relationship between
$\Delta\Omega$ and $\sqrt{\vert I_1I_2\vert}$.
At the quantum phase transition points
the spin chain becomes gapless
and $c$ depends linearly on $T$.
The ridges seen in Figs. 3a, 3b
(which correspond to maxima in the dependence $c$ vs. $\Omega$
as $T$ varies) single out the boundaries
of quantum critical regions \cite{002}.
These boundaries correspond to a relation  $\Delta\sim kT$
that can be checked
by comparison with the data for $\Delta$ vs. $\Omega$
reported in Figs. 1a, 1d.
As can be seen from Fig. 3
the $c(T)$ behaviour for $\Omega$
slightly above or below $\Omega_c$
changes crossing the boundaries of quantum critical regions.
Furthermore
we notice that for $\Delta\Omega=1.5$
an additional low-temperature peak appears in $c(T)$.

To end up,
let us turn to a chain of period 3
for which the analytical and numerical calculations
presented above can be repeated.
The quantum phase transition points follow from the condition
${\cal{A}}_6(0)=0$,
i.e.,
$\Omega_1\Omega_2\Omega_3=\pm I_1I_2I_3$.
Depending on the parameters
either two, four, or six
quantum phase transitions are possible.
This behaviour is illustrated in Fig. 4 for a chain of period 3
with
$\vert I_1\vert
=\vert I_2\vert
=\vert I_3\vert =1$,
$\Omega_{n}=\Omega+\Delta\Omega_n$,
$\Delta\Omega_1+\Delta\Omega_2+\Delta\Omega_3=0$.

To summarize,
we present exact results
for the thermodynamics
of a regularly alternating
$s=\frac{1}{2}$ Ising chain in a transverse field.
Furthermore,
for special parameter values also
the exact GS and spin correlation functions
are given.
For parameters
for which the correlation functions are not accessible
for rigorous analysis
we present exact numerical results
for finite but large systems.
From these numerical data
we can extract the order parameter and the correlation length
in high precision.
We find,
that the quantum phase transition present in the uniform chain
is not suppressed by deviation from uniformity
in kind of a regular alternation of bonds and fields.
On the contrary,
field alternation may lead to the appearance of additional transitions.
The number of phase transitions
for a given period of alternation
strongly depends
on the precise values of the parameters of the model.
For the gap $\Delta$,
the GS energy $e_0$,
the transverse magnetization $m_z$
and the static transverse susceptibility $\chi_z$
we can examine the critical behaviour exactly
and find the same critical indices
for the nonuniform and the uniform chain.
The presented study can be extended 
for examining the ground state properties 
of $s=\frac{1}{2}$ transverse Ising model 
on one-dimensional superlattices.

\vspace{5mm}

The present study was partly supported by the DFG
(projects 436 UKR 17/7/01 and 436 UKR 17/7/02).
O. D. acknowledges the kind hospitality of the Magdeburg University
in the summer of 2002.
O. D. and T. K. were supported
by the STCU under the project \#1673.

\vspace{10mm}

FIGURE CAPTIONS

\vspace{5mm}

Figure 1.
The GS properties
of the transverse Ising chain of period 2
with
$\vert I_1\vert=\vert I_2\vert=1$
and
$\Omega_{1,2}=\Omega\pm\Delta\Omega$ in dependence on the transverse field
$\Omega$:
Energy gap $\Delta$
(dotted curves) (a,d),
transverse magnetization $m^z$
(solid curves) (a,d),
static transverse susceptibility $\chi^z$
(dashed curves) (a,d),
longitudinal sublattice magnetizations $\vert m^x_j\vert$
(triangles and solid curves
correspond to the data for $N=20$ and $N=600$, respectively) (b,e),
and
correlation length $\xi^x$
(triangles and squares
correspond to the data for $N=300$ and $N=600$, respectively) (c,f) for
$\Delta\Omega=0.5$ (a,b,c) and
$\Delta\Omega=1.5$ (d,e,f).

\vspace{5mm}

Figure 2.
The GS properties
of the transverse Ising chain of period 2
with
$\vert I_1\vert=\vert I_2\vert=1$
and
$\Omega_{1,2}=\Omega\pm\Delta\Omega$ in dependence on the  transverse field
$\Omega$:
Longitudinal sublattice magnetizations $\vert m^x_j\vert$ (a)
and
correlation length $\xi^x$ (b)
for chain length of
$N=300$ (triangles),
$N=600$ (squares),
$N=900$ (circles) sites and
$\Delta\Omega=0.99$ (open symbols),
$\Delta\Omega=1.01$ (full symbols).

\vspace{5mm}

Figure 3.
The low-temperature dependence of the specific heat
for the transverse Ising chain of period 2
with
$\vert I_1\vert=\vert I_2\vert=1$
and
$\Omega_{1,2}=\Omega\pm\Delta\Omega$;
$\Delta\Omega=0.5$ (a),
$\Delta\Omega=1.5$ (b).

\vspace{5mm}

Figure 4.
Number of quantum phase transitions 
present in the transverse Ising chain of period 3
with $\vert I_1\vert =\vert I_2\vert =\vert I_3\vert =1$
and
$\Omega_{n}=\Omega+\Delta\Omega_n$,
$\Delta\Omega_1+\Delta\Omega_2+\Delta\Omega_3=0$.
The dark, grey, or light regions
correspond to parameters where the system exhibits
two, four, or six quantum phase transitions,
respectively.

\clearpage

\begin{figure}
\vspace{0mm}
\epsfxsize=120mm
\epsfbox{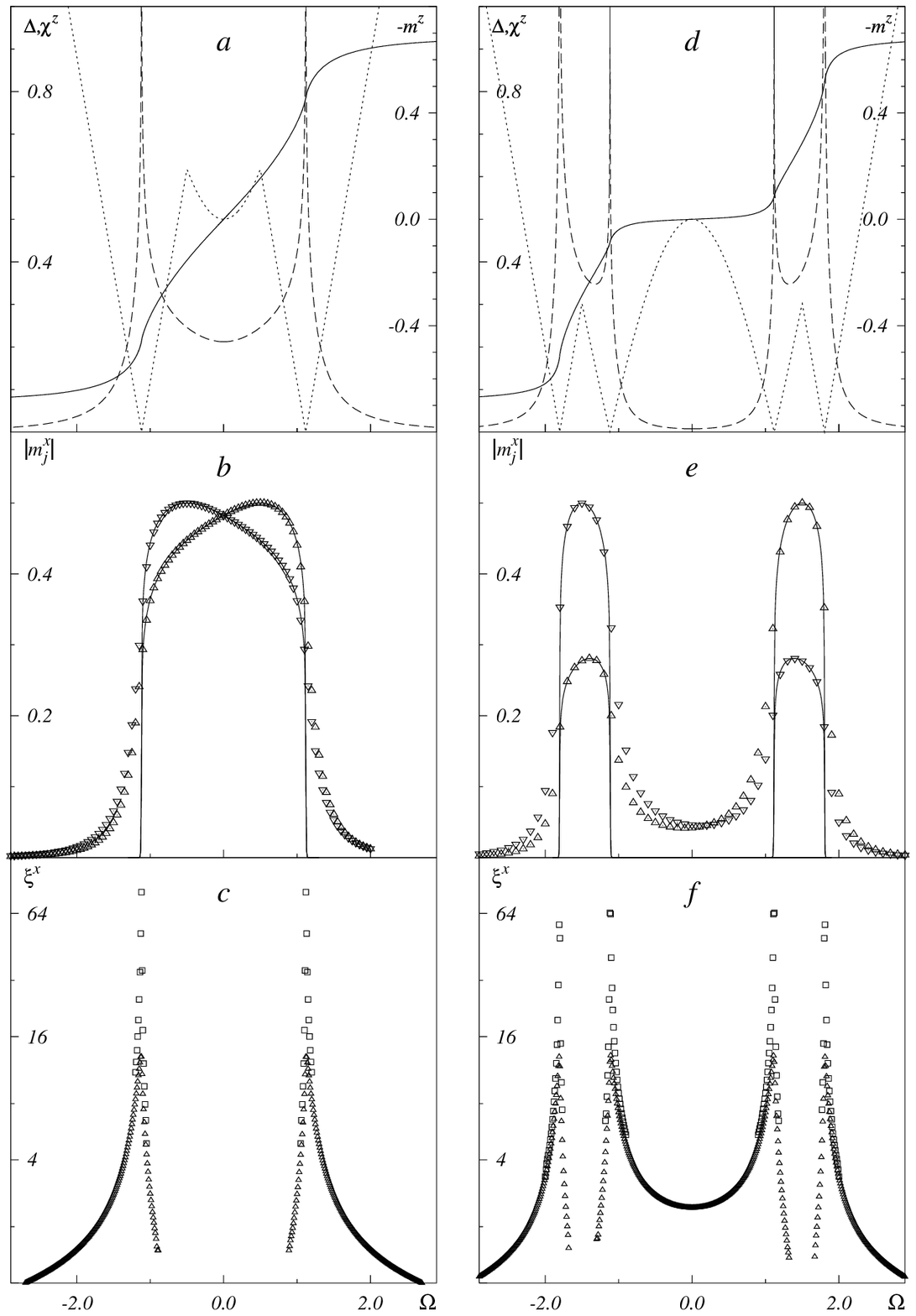}
\vspace{0mm}
\caption{\small{ }}
\end{figure}

\clearpage

\begin{figure}
\vspace{0mm}
\epsfxsize=70mm
\epsfbox{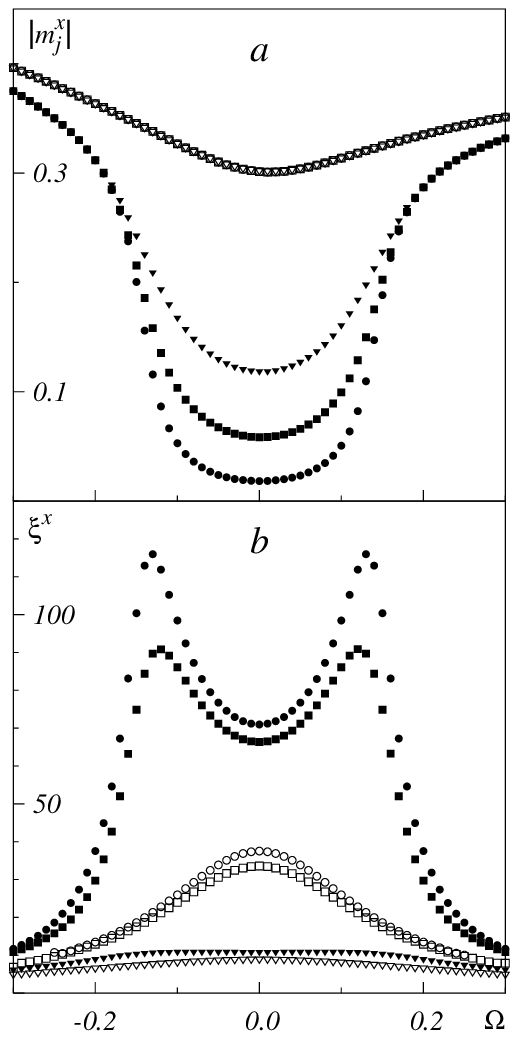}
\vspace{0mm}
\caption{\small{ }}
\end{figure}

\clearpage

\begin{figure}
\vspace{0mm}
\epsfxsize=180mm
\epsfbox{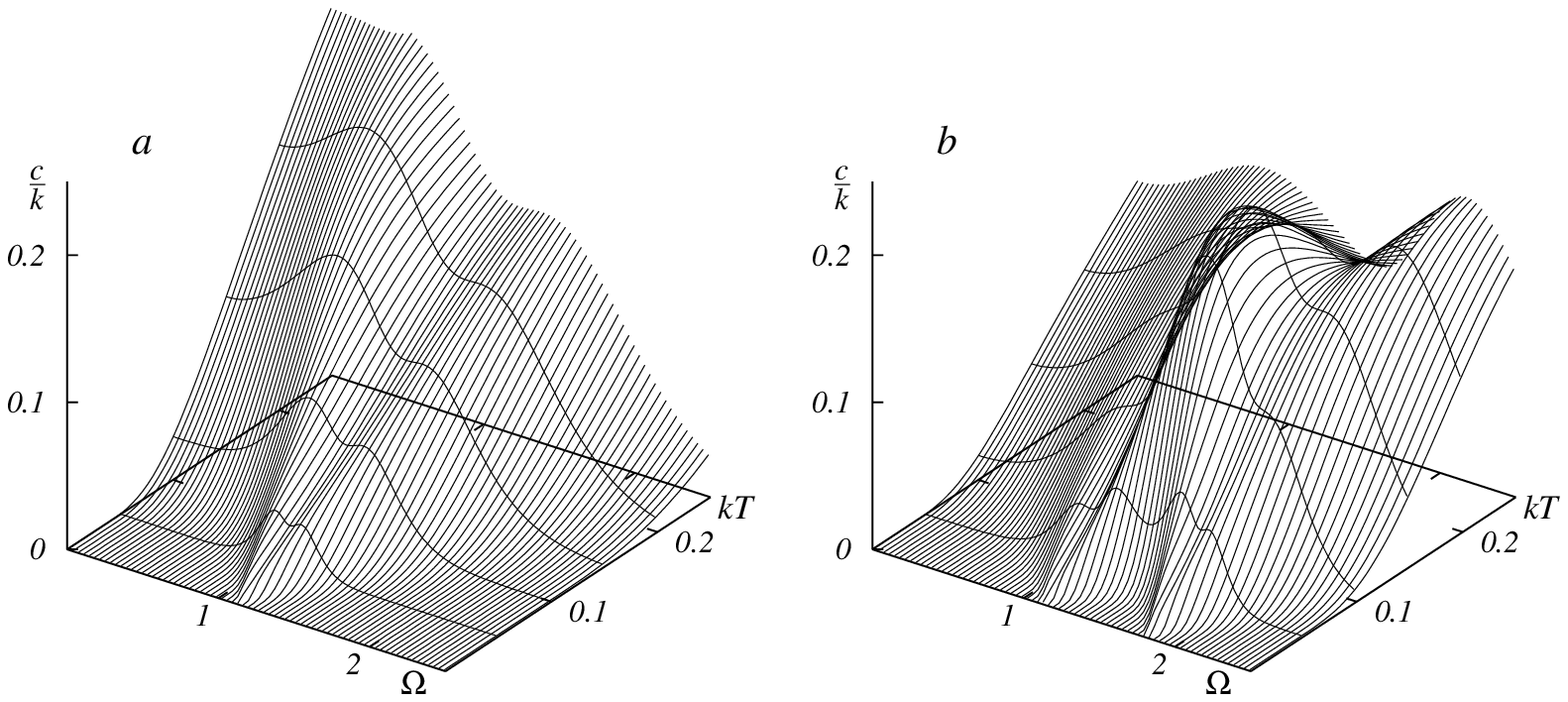}
\vspace{0mm}
\caption{\small{ }}
\end{figure}

\clearpage

\begin{figure}
\vspace{0mm}
\epsfxsize=80mm
\epsfbox{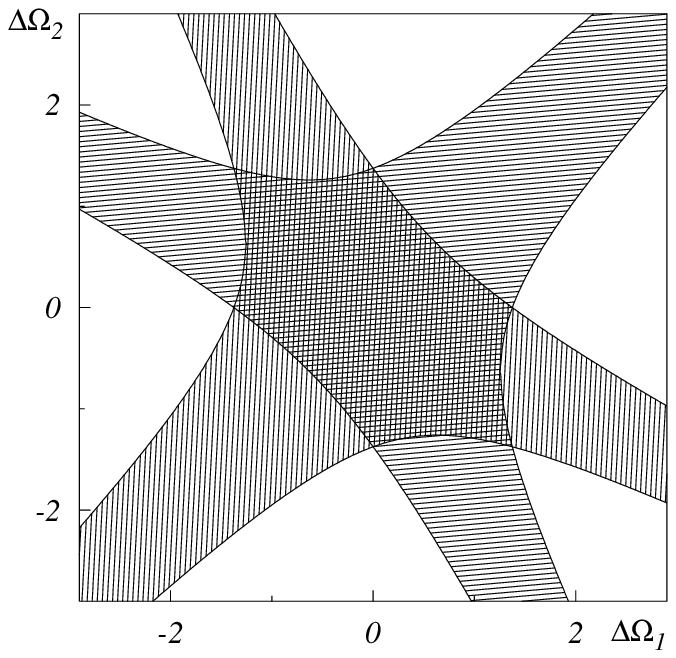}
\vspace{0mm}
\caption{\small{ }}
\end{figure}


\begin{thebibliography}{}
\bibitem{018}
Note,
that the critical condition found in Ref. \cite{017} 
(Eq. (6) of that paper)
in our notations reads
$\left(\Omega+\Delta\Omega\right)\left(\Omega-\Delta\Omega\right)
=I_1I_2$
and can yield only two critical fields.
However, by symmetry arguments
(i.e., performing simple rotations of spin axes)
one concludes 
that in addition to that condition 
there is an excitation of zero energy when 
$\left(\Omega+\Delta\Omega\right)\left(\Omega-\Delta\Omega\right)
=-I_1I_2$
and thus either two or four critical fields may appear.
\bibitem{019}
We may consider also a chain of period 2 with
$I_{1,2}=I\pm\Delta I$, $\Delta I\ge 0$
and examine the GS properties as $I$ varies.
Then we find
that for small
$\Delta I<\sqrt{\vert\Omega_1\Omega_2\vert}$
there are two values of
$I_c=\pm\sqrt{{\Delta I}^2 +\vert\Omega_1\Omega_2\vert}$,
at which the system becomes gapless,
whereas for large
$\Delta I>\sqrt{\vert\Omega_1\Omega_2\vert}$
there are four such values of
$I_c=\pm\sqrt{{\Delta I}^2\pm \vert\Omega_1\Omega_2\vert}$.
In general,
the condition (\ref{003})
may be tuned by some parameter(s)
influencing the local fields or/and exchange interactions.
\end{thebibliography}
\end{document}